\documentclass[twocolumn]{article}
\usepackage[a4paper, total={6in, 8in}]{geometry}
\usepackage{amsfonts}
\usepackage{amssymb}
\usepackage{amsmath} 
\usepackage{amsthm}
\usepackage{geometry}
\usepackage{graphicx}% Include figure files
\usepackage{enumitem} 
\usepackage{algorithm}
\usepackage{algcompatible}
\usepackage{algpseudocode}
\usepackage[colorlinks=true, allcolors=blue]{hyperref}
\usepackage[utf8]{inputenc} 
\usepackage{url}
\usepackage{cleveref}
\usepackage[english]{babel} 
\usepackage{subfiles}
\usepackage{xfrac}
\usepackage{multirow}
\usepackage{balance}
\usepackage{booktabs}
\usepackage{multicol}
\usepackage{graphicx} % Required for inserting images
\usepackage{tikz}
\usetikzlibrary{arrows.meta}
\usepackage{authblk}
\usepackage{natbib}
\usepackage{subcaption}
\providecommand{\keywords}[1]
{  \small	
  \textbf{\textit{Keywords---}} #1
}

\begin{document}

% POSSIBLE TITLES (ideally no more that 8 words) ranked longest to shortest
%  0. The Epistemic Limits of Empirical Finance: Causal Predation and Self-Reference
%  1. Causal Predation and Self-Reference: The Epistemic Limits of Empirical Finance
%  2. Epistemic Limits of Empirical Finance: Causal Predation and Self-Reference
%  3. Empirical Finance's Epistemic Limits: Causal Predation and Self-Reference
%  4. Epistemic Limits of Empirical Finance: Causality and Self-Reference
%  5. Epistemic Limits of Empirical Finance: Predation and Self-Reference
%  6. Epistemic Limits of Empirical Finance: Causal Self-Reference
%  7. Causal Predation and Self-Reference in Empirical Finance
%  8. The Epistemic Limits of Causation in Empirical Finance
%  9. Causal Predation and Self-Reference in Finance
% 10. The Epistemic Limits of Empirical Finance
% 11. The Epistemic Limits of Finance
\title{Epistemic Limits of Empirical Finance: Causal Reductionism and Self-Reference}
% \title{} 

%\author[*,1,2]{Daniel Polakow\thanks{Corresponding Author: \href{mailto:dpolakow@sun.ac.za}{dpolakow@sun.ac.za}}}
\author[1,2,*]{Daniel Polakow}
\author[3]{Tim Gebbie}
\author[2,4]{Emlyn Flint}
\affil[1]{Department of Statistics and Actuarial Science, University of Stellenbosch}
\affil[2]{School of Actuarial Science, University of Cape Town}
\affil[3]{Department of Statistical Sciences, University of Cape Town}
\affil[4]{Peresec, Cape Town}
\affil[*]{\small{Corresponding author: \href{mailto:dpolakow@sun.ac.za}{dpolakow@sun.ac.za}}}
%\date{November 2023}

\twocolumn[
\begin{@twocolumnfalse}
\maketitle
\textit{~~~~~~‘Everything that is born is necessarily born through the action of a cause’}
\begin{flushright}
Timaeus, Plato c. 360 BCE~~~~~~~
\end{flushright}
\textit{~~~~~~‘I can calculate the motion of heavenly bodies, but not the madness of people’}
\begin{flushright}
Sir Isaac Newton, 1720 CE~~~~~~~
\end{flushright}

\begin{abstract}
The clarion call for causal reduction in the study of capital markets is intensifying. However, in self-referencing and open systems such as capital markets, the idea of unidirectional causation (if applicable) may be limiting at best, and unstable or fallacious at worst. In this work, we critically assess the use of scientific deduction and causal inference within the study of empirical finance and financial econometrics. We then demonstrate the idea of competing causal chains using a toy model adapted from ecological predator/prey relationships. From this, we develop the alternative view that the study of empirical finance, and the risks contained therein, may be better appreciated once we admit that our current arsenal of quantitative finance tools may be limited to {\it ex post} causal inference under popular assumptions. Where these assumptions are challenged, for example in a recognizable reflexive context, the prescription of unidirectional causation proves deeply problematic.  
\end{abstract}
\vspace{0.4cm}

\vspace{1cm}

\end{@twocolumnfalse}]

\section{Introduction} 
%\section{Epistemic limits}
\label{sec:introduction}

There is an obvious resurgence of interest in causal inference. For good reason; causality is arguably the holy grail of any scientific enquiry \citep{Little1998} and has been deliberated since the time of \citet{Hume1739}. Existing coverage on causality in the philosophy literature is comprehensive and, for brevity, we will not be reviewing this in detail here.     

Rigor and a prescription of causal relevance has also pervaded finance and economics disciplines; including econometrics \citep{AP2010,CP2013,Ahmed2022} and investment management \citep{WilcoxGebbie2014,dePrado2023a}. Concurrently, Quantitative Finance (QF), which generically includes the investment and financial economics disciplines, is embarking on a purge of `incorrect' statistical methodologies. Such flawed statistical approaches have arguably resulted in a proliferation of false claims and charlatanism \citep{ZiliakMcCloskey2009,HL2014, BBBdePZ2014, Bailey2017, dePrado2015, dePrado2023a,dePrado2023b}. 

The uptake of defensible statistical methods is a progressive movement within QF that, along with more cautious distillation of causal pathways, supports scientific realism as well as positive economics.  QF might naturally benefit from such endeavors. However, while explaining economic phenomena is the intended result\footnote{{\it Sensu} the  positive economics of \citet{Friedman1953} and the apparent rise of economics as a `cyborg science' \citep{Mirowski2001-MIRMDE-2}.}, the movement also has the potential to incorrectly totalize disparate phenomena.  In particular, compared with prescriptive normative economics, such an approach can become uncoupled from market ontology, leading to a disconnect between the reality of capital markets and the assumptions -- and thus output knowledge -- of its reductionist studies.

In simpler terms, the assumption is that QF adheres to scientific laws, and hence is naturally only framed as ‘Science'.\footnote{This framing is epistemological. What is, or is not science is itself contested because of the sociological context of the activity \citep{Feyerabend1975}. The demarcation problem within QF resides in the dynamics of the system itself - and should not be confused with components that are driven by regulation and accounting practices (for example).}  In instances where this assumption fails and the paradigm becomes self-imposed, scientific methodology is unable to uniquely discern its own limitations or revise its usefulness.  As in the proverbial Emperor's New Clothes, we may be left with QF dressed in similar robes of causality. This could in part explain why financial crises are often inherent and unavoidable within a positive framing of QF.

Within most studies of QF, a single common assumption is pervasive; namely, that markets, being social systems, adhere sufficiently to epistemic norms. These epistemic norms carry with them {\it a priori} the necessity of unique, well-defined causal chains that can be meaningfully extracted from data in a positive economics framing and a reductionist scientific sense.

%\section{Epistemic Norms and Causal Chains}
\section{Epistemic norms and causal chains}

To understand epistemic norms in markets, first we note that there are some important assumptions underpinning economic thinking. Of course, there exist several different economic schools of thought. The dichotomy between neoclassical, or orthodox, and heterodox economics (arguably, everything else) is frequently made to smooth over and sometimes smother the pragmatic importance of these assumptions as a binary framing that casts idealisms and normative approaches in opposition to positivist perspectives.  

While orthodox approaches idealize an economic system in which the rationality of participants is key, heterodox economics emphasizes inefficiencies and imbalances generated by the system. Nonetheless, both modern orthodox and heterodox economic approaches aim to incorporate a range of imperfections within their standard frameworks to render them more realistic. In this setting, realism is defined as the ability to provide narrative explanations that are cohesive within a given ontology, or to make successful predictions.

At the heart of the positivist agenda, which resides within the broader scientific realist school of thought, is the goal of causal explanation; and unidirectional causation is the {\it de facto} supposition. In this regard, of contemporary relevance is the work spearheaded by \citeauthor{Pearl1995} and co-workers (\citet{Pearl1995,Pearl2009a,Pearl2009b}; \citet{ Pearletal2016}) who have formalized empirical causal modeling and hence laid the foundations of much of modern artificial intelligence. 

Unidirectional causality refers to $A$ causing $B$, for example the price action in $B$ being caused by market interest $A$, {\it ceteris paribus}.  In terms of the mathematics of causality \citep{Pearl1995}, we express $B$ as some function of $A$:
\begin{equation}
P[B=b \mid \mathrm{do}(A=a)] \neq P[B=b]. \label{eq:EQN1}
\end{equation}
The `do' operator sets $A=a$ by intervention. A diagrammatic representation of Eqn. [\ref{eq:EQN1}] is provided in Figure \ref{fig:CG1}.

Eqn. [\ref{eq:EQN1}] and its associated framework accommodates both neoclassical equilibria and heterodox dynamics, depending on the size and term of the arbitrageable phenomenon under consideration. Mathematician Robert \citeauthor{Buck1963} defined ‘self-frustrating’ predictions as forecasts that are initially true, but become false on public dissemination \citep{Buck1963}. Moreover, \citet{Popper1957} in his writings on social sciences refers to ‘self-limiting' predictions influencing the predicted event in a preventative sense. Markets are similarly understood to adapt as the outputs of scrutiny are assimilated.  This understanding is accommodated within the idealized neoclassic theory of economics.

Returning to Eqn. [\ref{eq:EQN1}], unidirectional causality does not permit $B$ to cause $A$ and it does not facilitate $A$ being limited by its subsequent impact on $B$ via self-limiting feedback. While causal graphs are typically linear and unidirectional, self-limiting processes with dampening feedback are commonly tolerated without being fatal to {\it ex post} inference. Importantly, contemporary QF is already suitably cognizant of this ‘adaptation’ of the market to new information \citep{dePrado2015, MP2016, Falkeetal2022}. 

More generally, adaptation is a feature common to `reflexive' systems; namely, systems that refer to themselves. While proponents of causal rigor recognize that markets are adaptive, the fact that they are, more broadly, self-referential is commonly overlooked. Specifically, markets may be possessed by dynamics that extend further than simply adaptation. 

A self-referential system is reflexive in nature. This is a surprisingly pervasive idea: the `Oedipus effect' of \citet{Popper1957}, the `self-fulfilling prediction' of \citet{Buck1963}, or the `back-coupling' of \citet{Morgenstern1972}. \citeauthor{MacKenzie2006} (of performativity literature renown) noted that the study of economics does more than simply describe, but rather shapes and changes the conditions of the economy, and society more broadly \citep{MacKenzie2006,MacKenzieetal2007}. It thus possesses the ability to analyze, describe, and modify its own structure, behavior, and properties. To say this phenomenology of self-reference is well-known, recurring and deeply studied is an understatement.\footnote{Self-referential study originates from \citeauthor{SpencerBrown1969}'s Laws of Form \citep{SpencerBrown1969} and extends to the calculus developed by \citet{Varela1975}. The field finds parallels in philosophy (radical constructivism; \citet{vonGlasersfeld1984}), control theory (e.g. second-order cybernetics; \citet{VonFoerster2003}), and economics (the Eigenform of markets; \citet{Kauffman2009}).  Daniel \citeauthor{HOF1999} discusses self-reference systems extensively in his popularist book \citep{HOF1999}.}  

\section{Muth's assumption}

The history and progression of economic thought is expansive, including the phenomenon of reflexivity. Reflexivity may pose a challenge to rational expectations theory and to the notion of an equilibrium.  The question of reflexivity has been given attention historically by many prestigious economists (including Nobel laureates Modigliani and Simon), yet it remains a topic of little serious interest in the field – for many of the reasons discussed. We do not review this literature further here but to note the following points. At the same time as \cite{ArrowDubreu1954}'s proof for the existence of a general equilibrium was published, two influential papers dealing with reflexivity were also published: \cite{GrunbergModigliani1954} and \cite{Simon1954}.  Both papers were a response to the unusual reflexivity predictions of \cite{Merton1948} - and both papers simply showed that \textit{‘it was possible’} (emphasis ours) that reflexivity did not matter in the accuracy of economic predictions. 

The seminal work of \cite{Muth1961} followed and was built upon the Grunberg-Modigliani-Simon proofs.  Muth’s work has endured and remains a standard characterization of rational expectations. Within it, reflexivity is assumed to have no impact - and this understanding was adopted as the only rational result. Somewhat ironically though, Modigliani and Simon were both known to have negative views on rational expectations (see \cite{Hands1990,Hands2013}).      

At different times, the implications of the Grunberg-Modigliani-Simon thinking resurfaces – for example the multiple ‘sunspot’ equilibria in the 80’s (\cite{Azariadis1981,CassShell1983}).  Conversely, while Grunberg-Modigliani-Simon has survived multiple attacks on its logic and mathematics, some authors consider the continuous nature of the Grunberg-Modigliani-Simon reaction functions to violate social understanding and conclude that Grunberg-Modigliani-Simon rarely ever applies (see \cite{Henshel1995}).

\section{Reflexivity in Capital Markets}

\citet{LWS2007} characterize two dynamics of reflexive systems: self-reinforcing and destabilizing (via positive feedback loop).\footnote{\citeauthor{Goodhart1984}'s law \citep{Goodhart1984}: {\it any observed statistical regularity will tend to collapse once pressure is placed upon it for control purposes}. \citet{CAMPBELL197967}'s law prior and the `Lucas Critique' \citep{LUCAS197619} are further examples of reflexive responses.} Markets are thus not limited to preventative feedback loops; instead, such loops may also be destabilizing.  For example, predictions such as fundamental market value can persist or change solely due to participants' beliefs.  Furthermore, runs on banks, asset bubbles, the volatility implied in option prices, booms-and-busts and the allure of factor exposures are all characteristic of a system possessed by self-referential properties.\footnote{Such examples may also, however, characterize systems that are not self-referencing.}.  

Examples of reflexive systems include the works of \cite{Merton1948}, \cite{Soros2009}, \cite{BrunnPedersen2009} and \cite{Bouchaud2024}.  Within such systems, sufficient conviction in an association can, in fact, result in ({\it i.e.} cause) that same association \citep{Polakow2010}. The notion of `spuriosity' thus also becomes ambiguous.   

Tony \citeauthor{Lawson2013}, the heterodox economist, explains why markets ultimately cannot adhere to epistemic norms. \citet{Lawson2013} defines closed systems to refer to a situation in which a correlation (an event regularity) occurs.  In the absence of conditions supporting such correlations, the system is then deemed to be ‘open’.  Markets, being a social system, are described as an ‘open system’.

%\begin{table*}[h!]
\textit{“… modern mainstream economics is awash with assumptions of perfect foresight or rational expectations, of efficient markets, and of market equilibrium, all, at least in the manner they are typically employed, essentially premised on the successful predictability of future outcomes; all supposing that in an open system actual outcomes can be effectively anticipated.  Indeed all such endeavour in effect treats open systems as if there are actually closed”} -- \citet{Lawson2013}.
%\end{table*} \label{tab:Lawsonquote}

Markets are unequivocally an open system.\footnote{Here, a closed system means a closed system of regular events, and implies predictability. An open system is then one in which prediction is itself not regularly achievable.} Such a system often involves recursive or circular relationships, where the system's components interact or define themselves in terms of other components within the same system. Of relevance here, self-referential systems are possessed of unusual properties, highly complex and challenging to understand or analyze, especially using traditional reductionist approaches and assumptions of causation \citep{Soros2013}. 

The dynamics of causal chains within self-referential systems are often paradoxical. \citet{Umpleby2010} notes that in self-referencing systems we transition from a classical paradigm of inferring A causes B, to higher-order dynamics with very different notions of stability, independence, structure and temporality. The very idea of model building and usage becomes nuanced. The presence of reflexivity will frequently result in circular reasoning -- deemed fallacious by contemporary academic standards.\footnote{Reflexivity further violates the {\it ad hominem} informal fallacy and the fallacy of accent: markets operate at both ‘observing’ and ‘participating’ levels \citep{Umpleby2010}.} The problem of causal emergence will frequently present itself in the form where multiple competing versions co-exist and compete \citep{Hoel2017}.

Markets comprise a system of interacting actors, with multiple levels of causation \citep{WilcoxGebbie2014}. The debate is not whether markets, being an obvious social system, are self-referential or not.  Rather, we believe the controversy revolves around two points.  Firstly, to what extent does such a self-referential system differ ontologically from the closed system we commonly assume? Secondly, can such a self-referential system produce behaviors and inefficiencies that are persistent and non-arbitrageable?  

Neoclassical economic thought, arguably the dominant paradigm, is premised on the basis that imbalances are only temporary and so the existence of short-term market imperfections are tolerated in the system.\footnote{In this vein was Soros’ view of reflexivity rebuffed by the economic establishment as ‘merely stating the obvious’ \citep{Soros2009}.}  

Heterodox thinking is more permissible to tolerating longer-term disequilibrium; {\it e.g.} simple ecological models applied to market dynamics result in endogenous inefficiencies and deviations from notional fundamental values \citep{GallaFarmer2013,Scholletal2021,SandersFarmerGalla2018}. \citet{Bouchaud2013} reviews modelling efforts that include heterogeneities as well as interactions, and proposes a potential unifying framework using Random Field Ising models for both endogenous ruptures and crises, including `self-referential' feedback loops for non-equilibrium phenomenology missing in classical economic models. \cite{HLoZ2023} examine the extent to which heterogeneity of investment styles can co-exist in the long-term, in contrast to the predictions wrought from traditional theory, and in response to social contagion.  The study builds on the Adaptive Markets Hypothesis \citep{Lo2004,Lo2017}, under which market participants behave (mostly) rationally using a set of natural-selection heuristics.  Such heuristics can become maladaptive during exogenous shocks.\footnote{The rise in popularity of dynamic stochastic general equilibrium (DSGE) models has been catalyzed by the need for policy analysis and forecasting away from classical economic core frameworks to permit for more richness and flexibility, but DSGE models have not been as successful as wished (see \cite{Stiglitz2018}).  We further note that a return to some notional equilibrium is an inherent feature in all DSGE models.}   

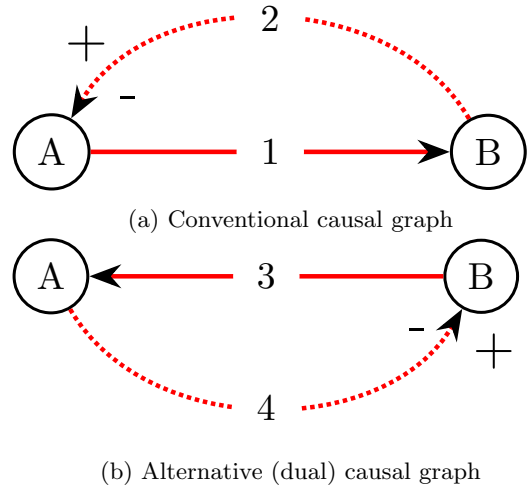
\begin{figure}[ht!]
\begin{center}
\captionsetup[subfigure]{justification=centering}
\begin{subfigure}[c]{0.45\textwidth}
     \resizebox{0.95\textwidth}{!}{
        \begin{tikzpicture}
\begin{scope}[every node/.style={circle,thick,draw}]
    \node (A) at (0,0) {A};
    \node (B) at (4,0) {B};
\end{scope}
\begin{scope}
    \node [anchor=west] (+) at (0,1) {\Large +};
    \node [anchor=west] (-) at (0.5,0.5) {\Large -};
\end{scope}
\begin{scope}[>={Stealth[black]},
              every node/.style={fill=white,circle},
              every edge/.style={draw=red,very thick}]
    \path [->] (A) edge node {$1$} (B);
%        \path [->] (B) edge[bend right=60] node {$2$} (A); 
\end{scope}

\begin{scope}[>={Stealth[black]},
 every node/.style={fill=white,circle},
              every edge/.style={draw=red,very thick, densely dotted}]
\path [->] (B) edge[bend right=60] node {$2$} (A); 
\end{scope}

\end{tikzpicture}
     }
\caption{Conventional causal graph} \label{fig:CG1}
\end{subfigure}
\captionsetup[subfigure]{justification=centering}
\begin{subfigure}[c]{0.45\textwidth}
   \resizebox{0.95\textwidth}{!}{
        \begin{tikzpicture}
\begin{scope}[every node/.style={circle,thick,draw}]
    \node (C) at (0,-1) {A};
    \node (D) at (4,-1) {B};
\end{scope}
\begin{scope}
    \node [anchor=west] (+) at (3.8,-1.7) {\Large +};
    \node [anchor=west] (-) at (3.2,-1.5) {\Large -};
\end{scope}
\begin{scope}[>={Stealth[black]},
              every node/.style={fill=white,circle},
              every edge/.style={draw=red,very thick}]
    \path [->] (D) edge node {$3$} (C);
    % \path [->] (C) edge[bend right=60] node {$4$} (D); 
\end{scope}

\begin{scope}[>={Stealth[black]},
 every node/.style={fill=white,circle},
              every edge/.style={draw=red,very thick, densely dotted}]
\path [->] (C) edge[bend right=60] node {$4$} (D); 
\end{scope}

\end{tikzpicture}
    }
\caption{Alternative (dual) causal graph} \label{fig:CG2}
\end{subfigure}
\end{center}
\caption{A conventional unidirectional causal graph of $A$ causing $B$ (path $1$) is shown in Fig. \ref{fig:CG1}; here with the addition of a negative (self-limiting feedback loop, denoted ``-"), and a positive (destabilizing feedback loop, denoted ``+"), both of the latter dependencies along path $2$. Fig. \ref{fig:CG1} (path 1 only) graphically represents Eqn. [\ref{eq:EQN1}]. This can be compared to the alternative unidirectional causal graph shown in Fig. \ref{fig:CG2} (path 3 only), which represents Eqn. [\ref{eq:EQN2}], where $B$ is shown causing $A$ along path $3$, with the addition of negative (self-limiting feedback loop, denoted ``-"); and positive (destabilizing feedback loop, denoted ``+"), the latter both along path 4. Feedback dynamics between these causal chains, \ref{fig:CG1} and \ref{fig:CG2}, may be indistinguishable unless intervention is involved.}
\label{fig:CausalGraph}
\end{figure}

Yet, it appears that both schools of thought -- neoclassical and heterodox -- retain the view that unidirectional causality remains relevant as a formalism, even if these linear cause-effect or circular dampening graphs exist only as component processes enmeshed in an ensemble of interactions of a more complex system \citep{Hardy2001,Hommes2021}.

When unidirectional causality breaks down, or becomes unstable, the models we utilize to predict economic phenomena will also naturally fail. \cite{Hendry1983, Hendry1985} studied predictive failures in econometric modelling, and developed models that assisted in identifying {\it ex post} structural breaks and regime shifts.  Hendry noted these models did not assume laboratory-like closed-systems and further permitted for changes within the parameterization of econometric models (the relevance thereof is discussed below).   \cite{Sornette2012} developed a self-exciting conditional Poisson Hawkes model that preemptively {\it ex ante} identifies when reflexivity results in `critical' levels of endogeneity - often thereafter resulting in market crashes.  

Financial systems are characterized by some degree of empirical continuity and stability, until such time as they are not, at which point our understanding and our models cease to function, at least temporarily, and in crisis.  As already alluded to, a positivist framing of QF models is unlikely to accommodate reflexivity (and the breakdown of unidirectional causation) before such financial crises.

If causality is in fact more malleable in a reflexive system (perhaps to the point of being wholly subjective) there is a competing reality that always requires admission as a hypothesis in any system prone to reflexivity, and that is: B causes A ({\it ceteris paribus}):\footnote{{\it e.g.} the price action in B causes the market interest A.}   
\begin{equation}
P[A=a \mid \mathrm{do}(B=b)] \neq P[A=a]. \label{eq:EQN2}
\end{equation}

A diagrammatic representation of {Eqn. [\ref{eq:EQN2}]} is provided in Figure \ref{fig:CG2}. It is critical to realize that the feedback dynamics embedded in Eqn. [\ref{eq:EQN1}] will effectively be indistinguishable from those of {Eqn. [\ref{eq:EQN2}]} unless intervention is invoked.\footnote{$B$ impacting $A$ via path 2 may be indistinguishable from $B$ causing $A$ via path 3, unless by control (intervention), and then only under certain conditions. This intervention is inevitably top-down in nature \citep{CraverBechtel2007,Aulettaetal2008}.} 

\citeauthor{Bouchaud2008} explains how disappointing quantitative success has been within the economic sciences, and motivates for a break away from classical economic tools \citep{Bouchaud2008,Bouchaud2013}. \cite{LWS2007} note that within economics, it is not possible to draw conclusions on the logical legitimacy of any reflexive mechanism. \cite{Shaikh2013} noted that even heterodox economics maintains the notion of an equilibrium and that different economic foundations are required for adequately dealing with reflexivity. The scientific fallibility of economics was identified and discussed comprehensively in the early 1980s \citep{McCloskey1983}. \citet{Soros2013} argued plainly that markets did not benefit from the same epistemic norms of science.  In particular, if financial models operate within a self-referential framework they can be non-causal, resulting in an inability to predict \citep{DeScheemaekere2009}.

In summary, the presence of reflexivity poses a clear challenge to the epistemology of QF since it is not accommodated within a unidirectional causality paradigm. The reason why reflexivity does so is simply stated: chains of causality are ostensibly malleable by the actors within markets. For guidance (and illustration) we turn to a well-understood set of complex systems: ecology.\footnote{Ecosystems are usually thermodynamically open, but causally closed with some regularity and predictability.}    

%\section{Modeling causal predation with self-referencing}
\section{Causal predation}

Extending ecology metaphors within finance has precedent \citep{Scholletal2021}. Contemporary empirical views provide examples of predator-prey models explaining, for example, S\&P500 dispersion in terms of contrarian and trend-following agents competing with and predating upon each other \citep{Lux2021}. As a useful metaphor, we explore the co-existing dynamics of two competing causal chains within a similar self-referential system, adapted from ecology \citep{May1976,Sugiharaetal2012}. This forms an illustrative toy model of simple reflexive causal chains that can generate complicated dynamics.\footnote{The concept of `regularity' can be further restricted -- {\it e.g.} Granger causality as the measure of predictability \citep{Sugiharaetal2012} -- this does not change our argument.}

Consider a single self-dependent and time-dependent feature $x(t)$ that can be well modeled by a logistics function, here with rate of change parameter $r$, and carrying capacity $K$
\begin{equation}
\dot{x}(t) = rx(t) (K-x(t)). \label{eq:LM}
\end{equation}
where the dot represents a time derivative. The feature $x$ could be taken to represent the number of individuals in a population, or the log-price of an asset, or the wealth associated with the accumulated profit-and-loss of some trading strategy conditioned on various information sources. This model can be randomly perturbed, although we keep the argument deterministic. 

Once discretized, the model demonstrates the well-understood bifurcation phenomena shown by \citet{May1976}. This is shown by replacing the feature derivative with the derivative of a logarithm of the feature from Eqn. [\ref{eq:LM}] and converting it into a difference equation: $\ln x_{n+1} - \ln x_n = r (K-x_n)$. By taking the exponential of the log-difference equation we find the Logistics Difference Equation (LDE)\footnote{The LDE is a rescaled form of the Ricker map \citep{Ricker1954}. Using the first-order expansion $e^z=1+z+O(z^2)$, Eqn. [\ref{eq:LDE}] gives $x_{t+1}\simeq(1+rK)x_t-rx_t^2$. Defining $R=1+rK$ and $u_t=(r/R)x_t$ gives $u_{t+1}\simeq R u_t(1-u_t)$, the standard logistic map studied by \citet{May1976}. Thus the logistic map follows after first-order expansion and rescaling, with logistic-map parameter $R=1+rK$.} :\begin{equation}
x_{n+1}=x_{n} e^{r(K-x_n)}. \label{eq:LDE}
\end{equation}

We now progress the metaphor, but for coupled features that include an interaction term. For simplicity we restrict ourselves to two features $x(t)$ and $y(t)$, with constant causal couplings\footnote{The Lotka-Volterra (LV) model has the form, for two chains $S_1$ and $S_2$: $\dot{S_1} = a S_1 - b S_1 S_2$, and $\dot{S_2} = c S_2 - d S_1 S_2$.}:
\begin{eqnarray}
    \dot{x}(t) &= r x(t) \left({K-x(t) - a_{yx} y(t)}\right), \nonumber \\ 
     \dot{y}(t) &= r y(t) \left({(K-y(t)- a_{xy} x(t)}\right).\label{eq:LDEs0}
\end{eqnarray}
There are two coupling parameters $a_{yx}$ and $a_{xy}$. If the two features were correlated then this parameter would be symmetric. The argument can be generalized to many such features.  To generate causal chains, we discretize the coupled differential equations \citep{Hastings1993,Sugiharaetal2012} and find the causal chain update equations:
\begin{eqnarray}
x_{n+1} &= x_n e^{r_x(K_x-x_n-a_{y \to x} y_{n})},\nonumber \\ % \label{eq:LDEx} \\ 
y_{n+1} &= y_n e^{r_y (K_y-y_n-a_{x \to y} x_{n})}. \label{eq:LDEs} 
\end{eqnarray}
This model has two different growth rates, two different carrying capacities, and two dependency variables. With $a_{y \to x}\neq0$, but $a_{x \to y}=0$ this would be unidirectional, from $y$ to $x$. These are now situated as part of a sequence of unidirectional causal relationships in a chain, where each iterate $x_n$, can be thought of as a realization of one node, and similarly, $y_n$ for the other, as a realization of the graph in Fig. \ref{fig:CausalGraph}.

\begin{figure}[ht!]
\includegraphics[width=0.54\textwidth]{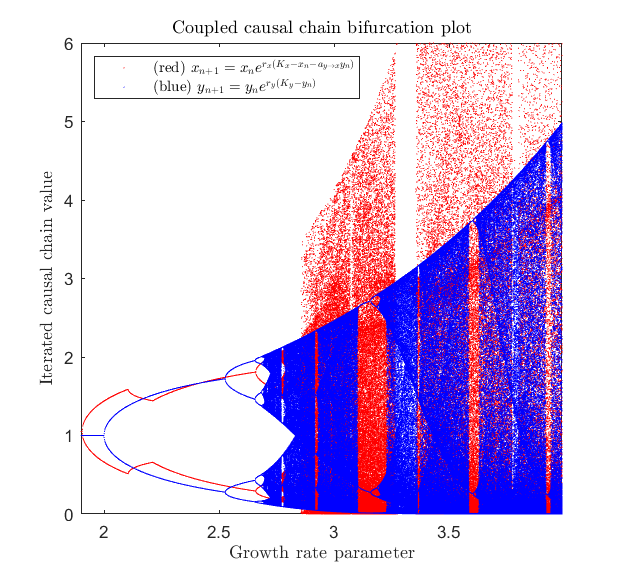}
\caption{A model of the iterative stability of sampled sequences of unidirectional causal graphs. This models the causal predation of the system using the Logistics Difference Equations (LDE) from Eqn. [\ref{eq:LDEs}]. The model includes self-interactions and cross-interactions where a feature $x$ has a unidirectional causal dependency on another feature $y$, but where $y$ is not dependent on $x$. The discretized features represent realizations from causal chains. The couplings introduce further instability and complexity. The y-axis is the feature value (population size, strategy price, or accumulated profit-and-loss) and is dependent on the growth factor $r$ given on x-axis, and the carrying capacity $K$. In this example $r_x= r$, $r_y = 0.95 r$, $K_x = 0.95$, $K_y=1.00$ and $a_{y \to x} =-0.1$; any variety of combinations can be chosen. The rounding significance is $\epsilon=10^{-4}$.}
\label{fig:coupledLDEgraph}
\end{figure}

We can simulate the chain of unidirectional causal graphs and visualize key features of the resulting dynamics by considering iterations of the coupled LDE. The resulting coupled bifurcation map is shown in Fig. \ref{fig:coupledLDEgraph}. This renders our key points concrete. 

We see that even if we explicitly know, with certainty, the deterministic and non-stochastic causal relationship between chains $x_n$ and $y_n$ (or even if we know the causal relationships between the underlying causal features $x(t)$ and $y(t)$) we cannot easily predict $x_n$ at all values of $r$ (nor $K$) even if $y_n$ is known with certainty itself.

As a simple representation for causal chain competition and predation, Eqn. [\ref{eq:LDEs}] and Fig. \ref{fig:coupledLDEgraph} have the further benefit of making clear the role of randomness.\footnote{Difference equations can be thought of as filters, this implies that erratic behavior due to sampling could be ameliorated using filtering {\it e.g.} low pass filtering.} For example, either $r$ (or $K$) can be changed to be uniformly distributed, and then sampled at exponentially distributed waiting times; perhaps representing exogenous shocks from top-down causal sources from the broader environment, either changing $r$, or $K$. We can also randomly change the interaction terms, and so on. These introductions of randomness do not improve the situation and can undermine attempts to close the system using additional filtering. In this framework, it is unreasonably optimistic to assume that one can meaningfully infer stable causal graphs uniquely from the data.

That is not to say causal graphs are not useful. In fact, for exploratory data analysis, or visualizing the hypothesized system state, or rendering assumptions transparent, causal graphs can be informative and should be encouraged. However, even if causal chains are unique and can be reliably extracted, this does not imply that the representation can faithfully render the dynamic properties of the system itself. 

In an open system, causality ultimately remains phenomenological. Markets are reflexive, adaptive and saturated with agents responding strategically to each other.  It is thus understood that coherence of perception around a common cause will often lead to forms of tight coupling with high cohesion, and thus fragility, being written into the financial system itself. 

\section{Conclusion}

Research may retrospectively support unidirectional chains of causality and their corresponding inference. However, the conditions in which {\it ex post} understanding can be extended to support {\it ex ante} prediction are limited. This is caused by the market's adaptation mechanism combined with the understanding that chains of causality are malleable by market actors. This necessitates a distinction of practical significance between ``model estimation", wherein the model is presumed to be normatively true and the data representative of the future, and ``model calibration", where one rather seeks to facilitate reliable decision-making under uncertainty using a model that may be misspecified or simply incorrect.\footnote{Here language is important, particularly with respect to its nuanced meaning in reference to its context. In much of the Finance and Econometrics literature there seems a reluctance to engage with epistemology to the extent that it challenges normative perspectives that are tied to both incentives and the historical development of ideas. For example, in some of the Econometrics and Economics literature the term model ‘identification’ is often used (\cite{KahnWhited2017}) to weaken the idea of model ‘estimation'. In reality, ‘identification' is still used to denote model selection within the estimation paradigm and, most importantly, is not used in the same way that model ‘calibration’ is used in the more pragmatic QF literature.} This theme -- a default necessity for calibration despite the presence of {\it ex post} confidence -- is well-known from derivative pricing \citep{Schoutensetal2004}. It is unlikely that these chains can be used without considerable risk for prediction or control in a forward-looking way. In financial economics, these risks may impact studies in asset pricing (where inputs are endogenous) more than, for example, studies within corporate finance (where inputs are often assumed exogenous).  That said, we are of the view that it does not serve the arguments to dilute the epistemology angle top-down, in lieu of a bottom-up argument from any specific economic or finance model.

If reflexive, the degree to which markets behave as a closed system is likely to be a function of only teleological conviction. This means that there may well be nothing tangible that supervenes various model representations. To the extent that certain QF actors dominate markets, a causal chain supporting their particular narrative will become established. The risk then exists of other actors benefiting from distinct causal chains coming to bear. The system is therefore prone to instability even in the absence of exogenous shocks. In the age of hyper-connectedness, the shifting of perception and sentiment has never been as fungible, or as inexpensive.\footnote{\cite{Baudrillard1981} called this the `precession of simulacra'.} It should be anticipated that without a more complete theory of complexity, QF will remain lost in a spiral of tautological self-reference.  

As scientists, and as academics of QF, we naturally aspire to high standards of critical thinking. Despite markets exhibiting unusual epistemic dynamics, scientific reductionism in QF is deployed as an inviolate prescription. However, the assumption of unidirectional causation will not always lead to the appropriate causal inference. In contrast, within a highly reflexive system, the pursuit of causality may, in fact, be doomed to failure.

Empirical finance may frequently find itself searching incorrectly for answers under the only epistemological light within which it can navigate. To the extent that markets closely proxy closed systems, the risk of violating directional causal assumptions is arguably limited. Yet, if markets do have the proclivity to ‘switch’ between different casual directions (as in Fig. \ref{fig:coupledLDEgraph}), QF is presented with an existential challenge. In such a scenario of overt reflexivity, empirical finance is expected to awaken within a deeply paradoxical associational relativity. Here, as in other contexts, realistic attempts at hypothesizing more complex causal models will become multifaceted and are less likely to give rise to useful inference \citep{Rasmussenetal2019,SaylorsTrafimow2020}.

None of this is novel in the real-world application of QF {\it i.e.} in option pricing, trading and asset management. One of the early approaches to dealing with the causation problem in factor investing -- path-dependencies and unidirectional dependencies -- was to introduce {\it ad hoc} models, such as those of \citet{HaugenBaker1996} and \citet{FersonHarvey1999}, and relax the linearity assumption of the pricing kernel \citep{WilcoxGebbie2015}.  This early modeling of dynamic temporal and multi-layered dependencies may appropriately be couched as attempts at ``machine learning" and are not essentially inventive, nor do they progress the central argument here. Hence, we deliberately avoid discussion of the ``data-science problem" or the merits or fallibility of any specific algorithm. The veracity of modeling tools in reflexive systems takes on a life of their own, and invariably has less to do with any computational tool or trick, and more to do with the nature of adoption and dissemination of the same. The fundamental nature of the problem remains, irrespective of what tools are fashionable.   

We believe causal and scientific thinking may ultimately be limited within the prescription of current QF tools for two reasons. The first presupposes that markets are not sufficiently reflexive to matter and reductionist approaches assuming unidirectional causality are valid. Within this paradigm, causal inference is limited to {\it ex-post}, unless the phenomenon under study is non-arbitrageable (because of the `tragedy of the commons’ \citep{Hardin1968}). The ability to forecast {\it ex ante} may ultimately be a rarity, and QF should be cognizant of overreach. Arguably, this understanding would accomplish much of the clean-up required within QF. The second limitation is that statistical and causal inference will plainly be applied incorrectly where reflexivity is present. For these reasons, we argue that empirical finance is not at risk of becoming a pathological science ({\it sensu} \cite{dePrado2015}). Rather, empirical finance is categorically a pathological science where it is able to act as one at all ({\it i.e.} where not sufficiently reflexive to matter), and patently not a science elsewhere.  

We close this section with words from one of the foremost thinkers in causal economics, Joshua Angrist who noted: \textit{`… Perhaps it’s worth restating an obvious point. Empirical evidence on any given causal effect is always local, derived from a particular time, place, and research design. Invocation of a superficially general structural framework does not make the underlying variation or setting more representative. Economic theory often suggests general principles, but extrapolation of causal effects to new settings is always speculative. Nevertheless, anyone who makes a living out of data analysis probably believes that heterogeneity is limited enough that the well-understood past can be informative about the future.'} \citep{AP2010}. Heterogeneity is certainly limited in some economic systems - for example where consumer preferences are simple and demand forecastable; similarly in commodity driven economies, single-currency zones and in traditional societies etc.  Yet, the challenge remains the surprising frequency with which the well-understood past does not in fact inform the future elsewhere. For example, it has been noted that the only consistent feature of economic and quantitative finance disciplines is the distinct inability to predict and avert crises \citep{Bouchaud2008}.   

The disparity between the two broad schools of economic thought is perhaps dialectical, being two sides of a causally confused coin. As yet, we do not have a viable alternative paradigm that accounts for the anomalies, and that can provide both a general and operational approach to the heterodox scientific project (\cite{Ghilarduccietal2023}). Such a viable alternative is a key requirement for any Kuhnian paradigm shift. In deeply reflexive systems, the heterodox project may in fact lack the necessary scientific foundation -  since it is without the predictive reach of causality.  This acknowledgement is belated to the discipline, and the perils of ignoring this phenomenological reality are potentially grave.  

\section{Beyond Regularities}

Causal predation (and the conditions of causal chain co-existence) are not satisfactorily assimilated in current thinking. It is clear that QF (and economics more broadly) would benefit from a better understanding of what knowledge is likely to be impacted by such reflexivity, and consequently the risks of attaching policy intervention, regulation or investment to the same.  This challenge is of vital importance, but remains epistemological and difficult.  Doing so entails a deeper understanding and questioning of model assumptions as well as entering into informal methods (such as using conceptual models and expert judgment) to better understand how and where working models might fail. 

\cite{Lawson1997} described a system that vacillates between predictability and non-predictability, as a contrastive demi-regularity.  Within periods of empirical continuity, predictable regularities do exist.  Outside of these periods, these predictable regularities break down, sometimes permanently.  Lawson proposes that in dealing with systems of demi-regularity, regularities are not deemed to be invariant, rather probabilistic and liable to change with underlying structure and mechanisms.  This then is the first prescription to any QF model validation process - accepting the potential impact of reflexivity and therefrom understanding potential fallibility in terms of (and expanding) what \cite{Derman1996} terms ‘Incorrect Model' model risk.  We note that reflexivity does not feature in any mainstream tertiary educational curricula or in professional designations, whereas it certainly should.

Second, it is important to aim to identify and understand the causal mechanisms generating the predictable phenomena and also then the conditions for their continuation or disintegration. For the purposes of this exercise, causal graphs ({\it sensu} \cite{Pearl1995}) are particularly useful as a point of departure, since they render causal assumptions explicit.  Third, model validation would consider out-of-the-box challenges to the stated causal assumptions, and their potential implications.  Lastly, an assessment of the impacts (including financial) of model failure (for any reason) should be presented.  Risk-management is acknowledged to be imperfect, as are the additions proposed here. Regardless, these provide a useful and unconventional stress-test in terms of formally acknowledging, to a greater degree, what may be normative in any QF model positivism.  

We look forward to the progression of future research, and methodological developments that are more duly cognizant of the impact and limitations of reflexivity.

\section{Acknowledgements}

We thank Diane Wilcox, Didier Sornettte, Michael Jennions and Peter Cotton for conversations and comments.  The views expressed here are solely those of the authors.  

\setlength{\bibsep}{0.0pt} % spacing between entries
\scriptsize{
% \small{
\bibliography{causalFailure}
}

%\end{multicols}
\end{document}